\documentstyle[amssymb,aps,epsfig,float,twocolumn]{revtex}
\restylefloat{figure}
\title{Magnetic Order-Disorder Transition in the Two-Dimensional
Spatially Anisotropic Heisenberg Model at Zero Temperature}
\author{D. Ihle$^a$, C. Schindelin$^b$, A.~Wei{\ss}e$^b$, 
and H.~Fehske$^{b}$}
\address{
  $^a$Institut f\"ur Theoretische Physik, Universit\"at Leipzig,
  D-04109 Leipzig, Germany\\
  $^b$Physikalisches Institut, Universit\"a{}t Bayreuth, 
  D-95440 Bayreuth, Germany\\
  {\rm (\today)}
  }
\address{~\parbox{14cm}{\rm
    \medskip
The ground-state properties of the spin-1/2 
antiferromagnetic Heisenberg model with spatially anisotropic 
couplings on a square lattice are investigated
by a spin-rotation-invariant Green's-function 
approach and by Lanczos diagonalizations on 
lattices up to 36 sites supplemented by 
finite-size scaling. We focus on the anisotropy-driven 
transition from the N\'eel state to a paramagnetic 
state with antiferromagnetic short-range order
and on the spatial dependence of spin correlation functions. 
Our principal result is that a rather sharp crossover 
in the magnetic behavior occurs at the 
coupling ratio $R_0\simeq 0.2$ ($R=J_y/J_x$).}} 
\begin{document}
\maketitle
Low-dimensional spin systems described by quantum 
antiferromagnetic (AFM) Heisenberg models have attracted 
increasing attention, in particular with respect to the
crossover from one to two 
dimensions~\cite{ST89,PSZ93,AGS94,FMYK98,San99}.
In this context, one basic problem is the influence of spatial 
anisotropy on the staggered magnetization $m$ in the
ground state of the two-dimensional (2D) spin-1/2 AFM Heisenberg model
\begin{equation}
  {\cal H} = \frac{J_x}{2}
\Big[ \sum_{\langle i,j\rangle_x}{\bf S}_i{\bf S}_j 
+R  \sum_{\langle i,j\rangle_y}{\bf S}_i{\bf S}_j\Big] \,.
\label{model}
\end{equation}
Here $R=J_y/J_x$ (throughout we set $J_x=1$), and $\langle i,j\rangle_{x,y}$ 
denote nearest neighbors (NN) along the $x$-, $y$-directions on a
square lattice. In the model (\ref{model}) there occurs a transition
from a long-range ordered (LRO) N\'eel antiferromagnet at $R=1$
($m=0.3074$~\cite{WY94}) to a spin liquid with pronounced
AFM short-range order (SRO) at $R=0$ ($m=0$).
The essential question is whether the critical coupling ratio 
$R_c$ for the order-disorder transition is zero or has 
a finite value. Whereas the ordinary spin-wave theory~\cite{ST89}
gives $R_c=0.034$ and the one-loop renormalization-group
analysis of an effective spatially anisotropic nonlinear 
sigma model results in $R_c\simeq 0.047$~\cite{NH96}, 
the RPA spin-wave theories~\cite{MSS92,REHDM97}
and the chain mean-field approaches~\cite{ST89,Sc96}
yield $R_c=0$. From the Pad\'e approximants to Ising expansions 
for $m$, $R_c\lesssim 0.02$ was suggested~\cite{AGS94}.
As stated by Affleck et al.~\cite{AGS94}, renormalization-group
arguments do not give a definite answer whether or not the system
orders for arbitrarily weak interchain coupling $J_y$.
Recently, Sandvik~\cite{San99} developed a multi-chain 
mean-field theory and presented quantum Monte Carlo (QMC)
results for an effective 3-chain model and for full 2D lattices
(at $R=0.1$ and $R=1$) which provide strong evidence for $R_c=0$.

Experimentally, in the quasi-1D antiferromagnets 
$\rm Sr_2CuO_3$ and $\rm Ca_2CuO_3$ a very small ordered 
moment and N\'eel temperature $T_N$ were observed and found 
to decrease with increasing anisotropy~\cite{Koea97}. 
Based on a detailed estimate of the exchange 
integrals, Rosner et al.~\cite{REHDM97} obtained 
ratios of the order of $R\sim 10^{-2}$ and concluded
that the previous theories overestimate both $m$ and $T_N$.

Motivated by this unsettled situation, in this Letter we examine 
the order-disorder transition and the spin correlation functions of
arbitrary range in the 2D spatially anisotropic Heisenberg model 
by an analytical approach based on the Green's-function 
projection method and by a finite-size scaling analysis
of exact diagonalization (ED) data.
Our results indicate a rather sharp crossover in the spatial
dependence of the spin correlation functions  
in the model~(\ref{model}) at the coupling 
ratio $R_0\simeq 0.2$. To this end, we extend 
the spin-rotation-invariant theory of 
Refs.~\cite{ST91,WI97}, which yields a good description of spin correlation 
functions in the isotropic model~($R=1$),
to the anisotropic case. 

\mbox{
To determine the dynamic spin susceptibility} 
$\chi^{+-}({\bf q},\omega)=
-\langle\langle S_{\bf q}^+;S_{-{\bf q}}^-\rangle\rangle_{\omega}$ 
by the projection method, we consider the two-time 
retarded matrix Green's function 
$\langle\langle {\bf A};{\bf A}^{\dagger}\rangle\rangle_{\omega}$ 
in a generalized mean-field approximation~\cite{WI97} 
\begin{equation} 
\langle\langle {\bf A};{\bf A}^{\dagger}\rangle\rangle_{\omega}=
[\omega-{\frak M}'{\frak M}^{-1}]^{-1}{\frak M}
\label{gf}
\end{equation}
with the moments ${\frak M}=\langle [{\bf A},{\bf A}^{\dagger}]\rangle$
and ${\frak M}'=\langle [i\dot{\bf A},{\bf A}^{\dagger}]\rangle$. 

We choose the two-operator basis  
${{\bf A}=(S_{\bf q}^+, i \dot{S}_{\bf q}^+)^T}$ 
and get
\begin{equation}
\chi^{+-}({\bf q},\omega)=-\frac{M^{(1)}_{\bf q}}{\omega^2
-\omega_{\bf q}^2}\,,
\label{chi}
\end{equation}
where $M_{\bf q}^{(1)}=\langle[i\dot{S}_{\bf q}^+,S_{\bf q}^-]\rangle $ 
is given by
\begin{equation}
M_{{\bf q}}^{(1)}=-4 C_{1,0}(1-\cos q_x)- 4 RC_{0,1} (1-\cos q_y) 
\label{mq}
\end{equation}
with $C_{\bf r}=\langle S_{\bf 0}^+S_{\bf r}^{-} \rangle\equiv C_{n,m}$, 
and ${\bf r}=n {\bf e_x} + m {\bf e_y}$. The N\'eel ordering in the 
model~(\ref{model}) is reflected by the closure of the spectrum gap at 
${\bf Q}=(\pi,\pi)$ and by the condensation of that mode.
Thus, at $T=0$, we have 
\begin{equation}
C_{\bf r}=\frac{1}{N}\sum_{{\bf q} (\neq {\bf Q})} 
\frac{M_{\bf q}^{(1)}}{2 \omega_{\bf q}} \mbox{e}^{i{\bf q}{\bf r}}
+ C \mbox{e}^{i{\bf Q}{\bf r}}\,.
\label{crwi}
\end{equation}
The condensation part $C$ results from~(\ref{crwi}) with 
${\bf r}=0$ employing the sum rule $C_{0,0}=\frac{1}{2}$.
Then the staggered magnetization is calculated as 
\begin{equation}
m^2=\frac{1}{N}\sum_{\bf r}\langle {\bf S}_{\bf 0} {\bf S}_{\bf r} \rangle
 \mbox{e}^{-i{\bf Q}{\bf r}}=\frac{3}{2} C\,.
\label{smwi}
\end{equation}
To obtain the spectrum $\omega_{\bf q}$ in the approximation 
$-\ddot{S}^+_{\bf q}=\omega_{\bf q}^2 S^+_{\bf q}$,
we take the site-representation and decouple the products 
of three spin operators in $-\ddot{S}^+_{i}$ along NN sequences, 
introducing vertex parameters in the spirit of the decoupling
scheme proposed by Shimahara and Takada~\cite{ST91}:
\begin{equation}
S_i^+S_j^+S_l^-=\alpha^{x,y}_1 \langle S_j^+S_l^- \rangle S_i^+
+\alpha_2   \langle S_i^+S_l^- \rangle S_j^+\,.
\label{dcoup}
\end{equation}
Here, $\alpha_1^x$ and $\alpha_1^y$ are attached to NN correlation functions
along the $x$- and $y$-directions, respectively, and $\alpha_2$ is 
associated with longer ranged correlation functions.
We obtain 
\begin{eqnarray}
\omega_{\bf q}^2&=&1+R^2+2\alpha_2(C_{2,0}+4RC_{1,1}+R^2C_{0,2})
\nonumber\\[0.1cm]
&&\hspace*{-0.4cm}-[1+2\alpha_1^xC_{1,0}+4R\alpha_1^y C_{0,1}\nonumber\\[0.1cm]
&&+2\alpha_2(C_{2,0}+2RC_{1,1})]\cos q_x
\nonumber\\[0.1cm]
&&\hspace*{-0.4cm}-R[R+4\alpha_1^xC_{1,0}
+2R\alpha_1^y C_{0,1}\nonumber\\[0.1cm]
&&+2\alpha_2(RC_{0,2}+2C_{1,1})]\cos q_y
\nonumber\\[0.1cm]
&&\hspace*{-0.4cm}
+2\alpha_1^xC_{1,0}\cos 2 q_x+2R^2\alpha_1^yC_{0,1}\cos 2 q_y
\nonumber\\[0.1cm]
&&\hspace*{-0.4cm}+4R(\alpha_1^xC_{1,0}+\alpha_1^yC_{0,1})\cos q_x\cos q_y\,.
\end{eqnarray}
Note that our scheme preserves the rotational symmetry in spin space, 
i.e., $\chi^{zz}({\bf q}, \omega)\equiv \chi({\bf q}, \omega)=\frac{1}{2}
\chi^{+-}({\bf q}, \omega)$. 

\begin{figure}[!htb]
  \epsfig{file= 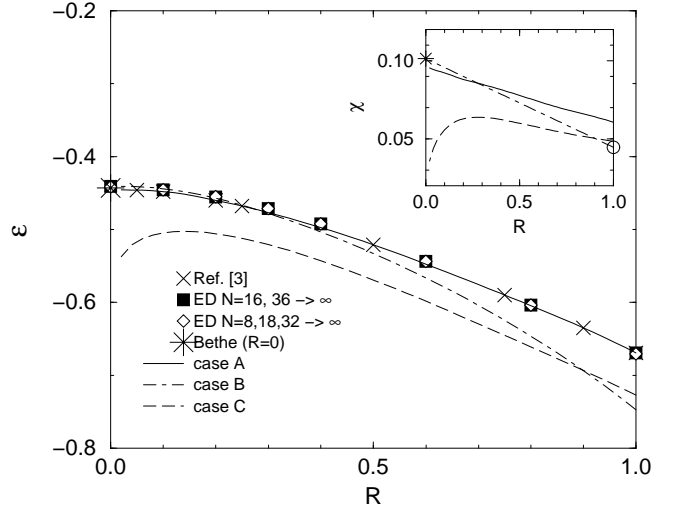, width = \linewidth}  
  \caption{$R$-dependence of the  ground-state energy per site $\varepsilon$ 
and the uniform static spin susceptibility $\chi$ (inset).}
\label{fig1}
\end{figure}

\begin{figure}[!htb]
 \epsfig{file= 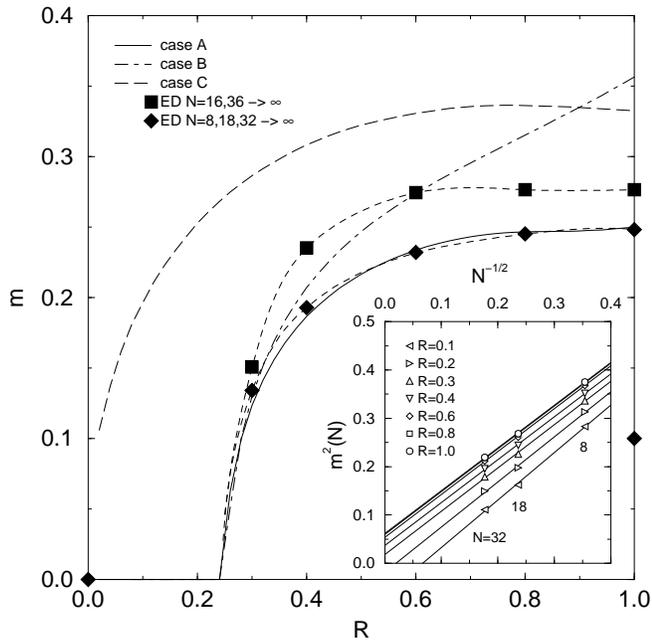, width = \linewidth}
 \caption{Staggered magnetization $m$ for different values of $R$. 
The short-dashed lines are fits to the ED data using  
$N=16$, 36 (squares) and tilted $N=8$, 16, 32 (diamonds) 
clusters with periodic boundary conditions, respectively. 
The inset demonstrates the deviation of the ED results  
from the $N^{-1/2}$ scaling law at $R\lesssim 0.3$
(solid lines are least-squares fits).}
\label{fig2}
\end{figure}

Considering the uniform static spin susceptibility  
$\chi=\lim_{{\bf q}\to 0} M_{\bf q}^{(1)}/2\omega_{\bf q}^2$,
the ratio of the anisotropic functions
$M_{\bf q}^{(1)}$ and $\omega_{\bf q}^2 = c_x^2 q_x^2 + c_y^2 q_y^2$ 
must be isotropic in the limit ${\bf q}\to 0$.
That is, the condition 
\begin{equation}
(c_y/c_x)^2=R C_{0,1}/C_{1,0}
\label{isocon}
\end{equation}
has to be fulfilled.
The theory has nine quantities to be determined self-consistently
(five correlation functions in $\omega_{\bf q}^2$, $m$, and three 
vertex parameters) and eight self-consistency equations 
(six Eqs.~(\ref{crwi}) including $C_{0,0}=1/2$, the LRO condition
$\omega_{\bf Q}=0$, and Eq.~(\ref{isocon})). If there is no LRO,
we have $\omega_{\bf Q}>0$, and the number of quantities and equations 
is reduced by one. As an additional condition we
treat three cases for the choice of the free $\alpha$ parameter:
(i) In {\it case A}, we adjust the ground-state energy per site 
$\varepsilon(R)$ taken from the Ising-expansion results 
by Affleck et al.~\cite{AGS94}
(Fig.\ref{fig1}). 
(ii) In {\it case B}, we fix the uniform susceptibility $\chi(R)$ 
by a linear interpolation between 
the exact value $\chi(0)=1/\pi^2$ and the QMC result 
$\chi(1)=0.0446$~\cite{Ru92} (star and open circle in the inset of 
Fig.\ref{fig1}, respectively), 
which is justified by the ED data of Ref.~\cite{PSZ93}.
(iii) In {\it case C}, we fit the free parameter to the QMC results
for $m(R)$ of the 3-chain mean-field theory by Sandvik~\cite{San99}
(Fig.\ref{fig2}).

Figure~\ref{fig1} compares the ground-state energy $\varepsilon(R)$ 
obtained by different approaches. Here the ED data is extrapolated 
using the scaling law $\varepsilon(N)-\varepsilon(\infty)\propto N^{-(R+2)/2}$, 
with an effective exponent agreeing with the known
values in the 1D~\cite{Ha86} and 2D~\cite{NZ89} cases 
(note that the 2D scaling exponent $-3/2$ seems to 
be more appropriate down to $R\simeq 0.25$).
In order to reduce cluster-shape effects, we use two different groups
of clusters of equal symmetry. Our ED data agrees with the 
Ising-expansion results by Affleck et al.~\cite{AGS94} and, in particular,  
with the Bethe value $\varepsilon(0)=-0.4431$. 
In the region $R\lesssim 0.25$, the ground-state energy of case~B
approximately reproduces the exact data. Equivalently, at
$R\lesssim 0.25$ the susceptibilities $\chi(R)$ in the cases~A and~B
nearly coincide (see inset of Fig.\ref{fig1}).

Our results for the order parameter $m$ plotted
in Fig.\ref{fig2} indicate an order-disorder 
transition at the critical ratio $R_c\simeq 0.24$
(note that the cases~A and~B give very similar results for 
$R\lesssim 0.25$, cf. also Fig.\ref{fig1}). The linear decrease
of $m$ with $R$ in the region $0.6 \lesssim R\lesssim 1$ in 
case~B is ascribed to the crude (linear) interpolation of
$\chi(R)$ in that case. To analyze the ED data 
for $m^2(N)$ calculated by Eq.~(\ref{smwi}), we use the 
finite-size scaling arguments given in Refs.~\cite{NZ89}
and~\cite{SZP96} for the 2D (frustrated) Heisenberg antiferromagnet 
and fit the data to the scaling relation 
$m^2(N)-m^2(\infty)\propto N^{-1/2}$
(inset of Fig.\ref{fig2}). Thereby, as noted in Ref.~\cite{SZP96},
the extrapolated $(N\to\infty)$ values for $m$ slightly depend on the
factor in front of the sum in Eq.~(\ref{smwi}): $N^{-1}$ vs. $(N+2)^{-1}$.
The finite-size extrapolation of $m$ for $N=16$, 36 depicted in
Fig.\ref{fig1} ($1/N$ prefactor) and calculated with an
$1/(N+2)$ prefactor agree, at $R=1$, with Ref.~\cite{SZP96}.
We find a transition to a spin-liquid phase at $R_c\simeq 0.24$ 
or $R_c\simeq 0.18$ depending on the chosen prefactor.
Let us emphasize the  coincidence of the
critical coupling ratios obtained in cases~A and B, and 
by Lanczos diagonalizations, where the critical ratio 
$R_c$ turns out to be nearly identical for both cluster 
sequences. 
To illustrate the finite-size scaling in more detail, 
in the inset of Fig.\ref{fig2} we show the least-squares fits of 
our ED results. The $N^{-1/2}$ scaling~\cite{NZ89}, when applied 
to the small systems (up to $6\times 6$), breaks down for $R\lesssim 0.25$.
As recently pointed out by Sandvik~\cite{San99,San99p}, at 
$R\lesssim 0.2$ the QMC simulations reveal an anomalous (non-monotonic) 
scaling behavior for square lattices, 
where very large systems are needed to access
the $N^{-1/2}$ law indicative of N\'{e}el order. 

To capture the behavior of $m(R)$ with $R_c=0$, let us consider
case~C. The results for $\varepsilon(R)$ and $\chi(R)$ shown in Fig.\ref{fig1}
strongly deviate, in the physically most interesting region of $R$, 
from those obtained in cases~A and~B. 
In particular, at $R\sim 0.2$, $\varepsilon(R)$ and
$\chi(R)$ exhibit a strange maximum.

\begin{figure}[!htb]
  \epsfig{file= 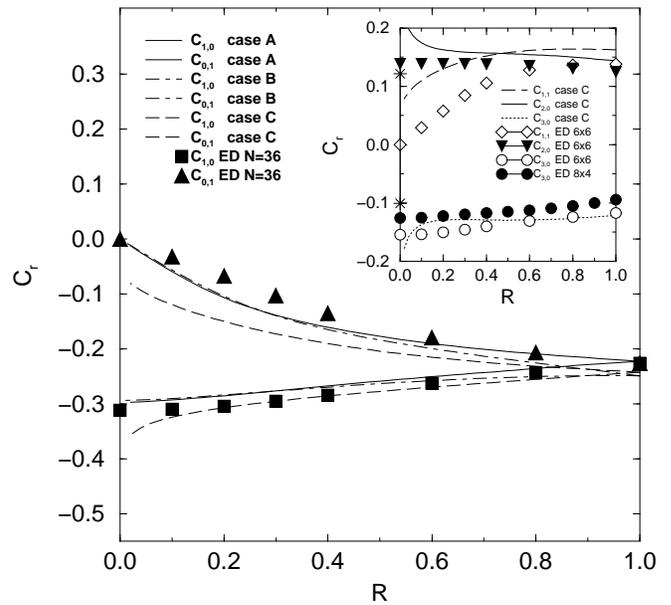, width = \linewidth}
  \caption{NN and longer ranged (inset) spin-correlation functions versus $R$.
Theoretical results are compared with ED data obtained for different
cluster sizes/shapes. In the inset, at $R=0$, 
stars denote the extrapolated values of the ED data 
for chains up to 32 sites using the scaling 
relation $C_{n,0}(N)-C_{n,0}(\infty)\propto N^{-2}$.}
\label{fig3}
\end{figure}

Our findings indicate a rather sharp change in the 
magnetic properties at $R_0\simeq 0.2$. 
Note that Affleck et al.~\cite{AGS94} found a poor convergence of
their Pad\'e approximants just below $R\lesssim 0.2$.
To shed more light into the crossover at $R_0$, we calculate
the dependence on $R$ of the two-spin correlation functions depicted
in Fig.\ref{fig3}.
The NN correlation functions in cases~A and~B reasonably agree with the
ED data for a $6\times 6$ system, whereas $C_{0,1}$ in case C notably
deviates from those results for $R\lesssim 0.25$. That is, in the $R$ region, 
where a finite magnetization is incorporated (case~C) as compared to
the $m=0$ solution in the cases~A and~B, the SRO is described less well.
The same qualitative behavior is found for the correlation functions
$C_{1,1}$ and $C_{2,0}$ (see inset of Fig.\ref{fig3}): At $R\lesssim 0.25$,
the results in cases~A and~B are in better agreement with ED data 
than the curve in case~C.
Note that the ratio $C_{0,1}/C_{1,0}$ determining the anisotropy in the
spin-wave velocities according to Eq.~(\ref{isocon}) 
does not show a decoupling 
transition ($C_{0,1}/C_{1,0}=0$) at a finite $R$ value, 
contrary to the suggestion by Parola et al.~\cite{PSZ93}.

From our results we suggest the following characteristics of the crossover
at $R_0\simeq 0.2$. In our self-consistent Green's-function theory 
the short ranged magnetic correlations,
in particular the fluctuations of the correlation functions about their
LRO limit, are described rather well. On the other hand, the suppression
of LRO by quantum spin fluctuations is overestimated. 
Moreover, if the correlation functions decrease slowly 
(subexponentially) towards their long-range limit,
our theory does not capture this behavior. 
Equally, the ED data for the small systems cannot be described by the
LRO scaling law $N^{-1/2}$. Our results indicate that this happens for
$R<R_0\simeq 0.2$.
Contrary, at $R>R_0$ the decrease of the correlation functions with
distance is sufficiently strong so that the LRO limit is nearly
reached at relatively short distances. This yields a reasonable scaling
behavior of our ED data (inset of Fig.\ref{fig2}). To sum up,
our results indicate that the crossover at $R_0\simeq 0.2$ may be 
accompanied by a rather sharp change in the spatial
dependence of spin correlation functions from strong to weak decrease
with distance. At $R<R_0$, the system starts to sense the 1D behavior,
characterized by an algebraic decrease of the correlation functions.

\begin{figure}[!htb]
  \epsfig{file= 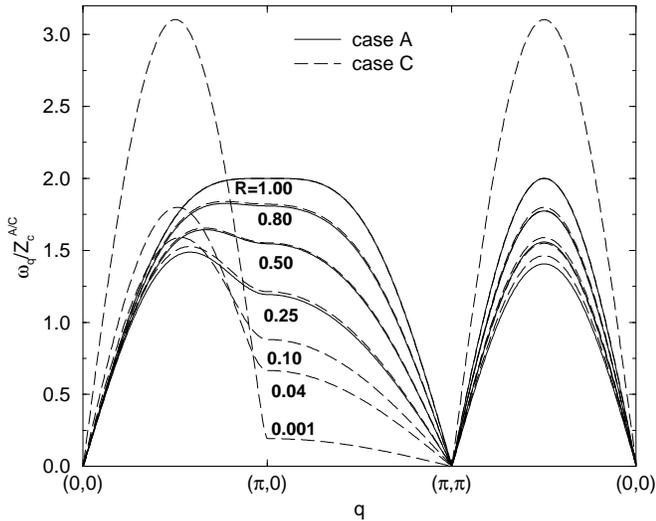, width = \linewidth}
  \caption{Spin-wave dispersion $\omega_{\bf q}$ of the 2D spatially 
anisotropic AFM Heisenberg model along the major symmetry directions
of the Brillouin zone.}
\label{fig4}
\end{figure}

Finally, in Fig.\ref{fig4} we plot the spin-wave spectrum in the ordered
state for the cases~A ($R>R_0\simeq 0.24$) and~C ($R>0$) scaled by 
$Z_c=c_x(R=1)/\sqrt{2}$, where $Z_c^A=1.36$ and $Z_c^C=1.58$. 
For $R=1$, in both cases we have $\omega_{\bf q}^2=4Z_c^2(1-\gamma_{\bf q}^2)$ 
with $\gamma_{\bf q}=(\cos q_x +\cos q_y)/2$.
With decreasing $R$, the anisotropy in $\omega_{\bf q}$ develops
gradually, where in case~C the 1D limit 
$\omega_{q_x}=2\sqrt{-\alpha_1^x C_{1,0}}|\sin q_x|$ is approached.
However, compared with the exact Bethe-ansatz result 
$\omega_{q_x}=\frac{\pi}{2}|\sin q_x|$, the spin-wave energy at $q_x=\pi/2$ 
is too high by a factor of about three.

To conclude, the strong evidence for the sharp crossover at the interchain
coupling ratio $R_0\simeq 0.2$ found by  analytical and numerical methods
should be corroborated by further QMC studies, especially on the distance
dependence of spin correlation functions. This is of particular interest
in developing appropriate theories to explain the magnetism in quasi-1D
quantum spin systems~\cite{REHDM97,Koea97}.

We thank S.-L.~Drechsler, R.~Hayn, W.~Janke, A. Kl\"umper,
J.~Stolze, and J.~Schliemann for useful discussions. 
Particularly we are indebted to A.~W.~Sandvik
for putting his (unpublished) QMC data to our proposal. 
H.~F. acknowledges the hospitality 
at the Universit\"at Leipzig, granted by the 
graduate college ``Quantum Field Theory''. The ED calculations
were performed at the LRZ M\"unchen, HLRZ J\"ulich, and the
HLR Stuttgart.
\vspace*{-.2cm}
\bibliography{ref}
\bibliographystyle{phys}
\end{document}